\documentclass{article}
\usepackage{spconf,amsmath,graphicx,hyperref}
\usepackage{multirow}
\usepackage{booktabs}
\usepackage{float}
\usepackage{subfig}

\title{The SJTU X-LANCE Lab System for MSR Challenge 2025}
\name{Jinxuan Zhu\textsuperscript{1}, Hao Qiu\textsuperscript{1}, Haina Zhu\textsuperscript{1, 3}, Jianwei Yu\textsuperscript{2}, Kai Yu\textsuperscript{1}, Xie Chen\textsuperscript{1, 3*} \thanks{* Corresponding author.}}

\address{
    \textsuperscript{1} X-LANCE Lab, School of Computer Science, MoE Key Lab of Artificial Intelligence,  \\ Jiangsu Key Lab of Language Computing, Shanghai Jiao Tong University 
    \textsuperscript{2} Microsoft Research
    \textsuperscript{3} SII \\ 
    \{zhujinxuan, chenxie95\}@sjtu.edu.cn
  }

\begin{document}
%
\maketitle
\begin{abstract}
This report describes the system submitted to the music source restoration (MSR) Challenge 2025. Our approach is composed of sequential BS-RoFormers, each dealing with a single task including music source separation (MSS), denoise and dereverb. To support 8 instruments given in the task, we utilize pretrained checkpoints from MSS community and finetune the MSS model with several training schemes, including (1) mixing and cleaning of datasets; (2) random mixture of music pieces for data augmentation; (3) scale-up of audio length. Our system achieved the first rank in all three subjective and three objective evaluation metrics, including an MMSNR score of 4.4623 and an FAD score of 0.1988. We have open-sourced all the code and checkpoints at \url{https://github.com/ModistAndrew/xlance-msr}.
\end{abstract}
\section{Introduction}

Recent advancements in music source separation (MSS) have achieved remarkable performance. However, a significant gap persists between academic research and practical audio engineering, as MSS models typically assume mixtures are linear combinations of clean sources, disregarding real-world signal degradations. The music source restoration (MSR) task \cite{msr} aims to bridge this gap by not only separating instruments but also restoring them to a state prior to any mastering artifacts (e.g., EQ, compression, reverb, codec), enabling applications like professional remixing and historical recording enhancement.

The MSR Challenge 2025 \cite{msrresult} poses a complex many-to-one mapping problem, which requires recovering unprocessed source signals from fully mastered and degraded mixtures across 8 instrument categories (\textit{vocals, guitars, keyboards, bass, synthesizers, drums, percussions, orchestral elements}). To address the scarcity of original, unprocessed audio for training, the challenge organizers have provided the training dataset RawStems in \cite{msr} and the validation dataset MSRBench \cite{msrbench}, enabling systematic research on MSR approaches. In response to this challenge, we propose a sequential BS-RoFormer \cite{bsroformer} framework that decomposes the MSR task into specialized modules for denoise, MSS, and dereverb. By leveraging pretrained models and implementing tailored training strategies, our system aims to achieve robust MSR performance.


\section{Methodology}
\label{sec:method}

\begin{figure}[htb]

\begin{minipage}[b]{1.0\linewidth}
  \centering
   \centerline{\includegraphics[width=8.5cm]{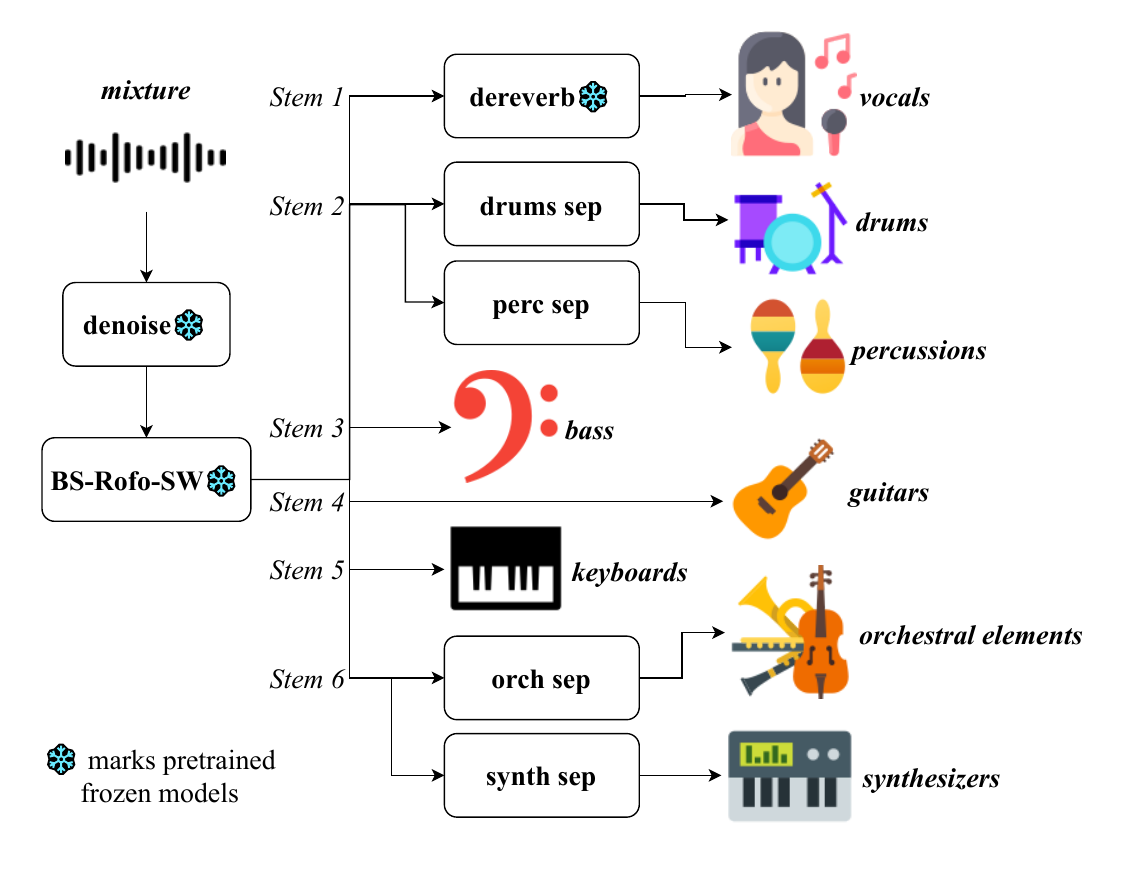}}
\end{minipage}
\caption{System architecture of the sequential BS-RoFormer framework. The input mixture is first processed by a denoise module to remove noise that adversely affects later separation process. The cleaned signal then passes through a frozen pretrained 6-stem separation model, followed by additional fine-tuned separation models for handling certain instruments. Finally, only the vocals stem undergoes dereverb.}
\label{fig:res}
\end{figure}

\subsection{BS-RoFormer}

BS-RoFormer is the currently state-of-the-art MSS model proposed by Lu et al. \cite{bsroformer}. It employs a band-split module to decompose the input complex spectrogram into subband-level representations, followed by Transformers that jointly model inner-band and inter-band sequences for multi-band mask estimation. We adopt the open-source implementation of BS-RoFormer from \cite{msst}. To maximize separation performance for each instrument, we train individual BS-RoFormer models per target stem rather than sharing Transformer layers across instruments.

\subsection{Separation model}

Separating eight instrument categories from a mastered mixture poses a significant challenge. We build upon a strong pretrained 6-stem BS-RoFormer model called BS-Rofo-SW (available at \url{https://huggingface.co/jarredou/BS-ROFO-SW-Fixed}) that separates \textit{bass, drums, other, vocals, guitars, piano (keyboards)}. To adapt to the target 8-stem setting, we train four new models dedicated to further decomposing the \textit{drums} and \textit{other} outputs into the finer categories of \textit{synthesizers, drums, percussions, and orchestral elements}, utilizing Transformer parameters of BS-Rofo-SW as initialization. The final result is a cascaded architecture with mixture first passing through the frozen 6-stem separator, followed by the fine-tuned models for stem refinement.

\subsection{Restoration model}

Restoration aims to recover real-world degradations present in the mixture. We focus on two most perceptually significant degradations: noise and reverb. Specifically, we employ (1) a denoise module applied first, since downstream separation models are sensitive to such broadband noise, and (2) a dereverb module applied only to vocals at the final stage, since reverberation for instrumental stems may be part of the intended timbre. The denoise-separate-dereverb sequence can be viewed as the inverse of the typical forward process of music recording, mixing, and effect application. We directly utilize checkpoints publicly available in the community \cite{msst}. Both denoise and dereverb models are Mel-band RoFormer models \cite{mel}, an updated version of BS-RoFormer. In practice, the dereverb module is bypassed if the output \(\text{Level} = 20 \log_{10}(\text{RMS}(x))\) drops by more than 10 dB, preventing removal of reverb-like vocal components (e.g., backing vocals).

\subsection{Training schemes}

\textbf{Data Cleaning and Mixing:} We combine the RawStems dataset with the MoisesDB dataset \cite{moises}, which provides additional music pieces with fine-grained instrument labels. RawStems is filtered to remove incorrectly labeled samples.

\textbf{Random Mixture Augmentation:} During training, we may randomly mix stems from different songs to create synthetic mixtures, increasing the diversity of training data.

\textbf{Scale-up of Data Length:} To enhance long-context modeling, we may extend training segments to longer durations (up to 10s), allowing the model to better capture temporal structures.

\section{Evaluation}
\label{sec:exp}

The model setting of BS-RoFormer mainly follows BS-Rofo-SW. Pretrained models operate on 44.1 kHz audio, while the data provided is 48 kHz, so resampling is necessary. We employ a combined loss of L1 loss and multi-resolution STFT loss. All models were trained on NVIDIA H200 GPUs with batch size 4. Training steps varied slightly but all exceeded 200k steps to ensure convergence.

To determine the optimal training scheme, we evaluated three configurations on the DT0 subset of MSRBench: Baseline (3s segments), Large (10s segments), and Large+Random (10s segments with random mixture augmentation). Performance differences were marginal across configurations. Based on the perceptually relevant FAD metric, we selected Large configuration for synthesizers, percussion, and orchestral elements, and Large+Random for drums for our final ensemble. Ablation studies of the pipeline were also carried out and can be reproduced from our repository.

\begin{table}[htb]
\centering
\caption{MSR Challenge 2025 results. Our system (xlancelab) achieved the best scores across all objective and subjective metrics. See \cite{msrresult} for details. }
\label{tab:compact_results}
\small
\begin{tabular}{lccc}
\toprule
\multicolumn{4}{c}{\textbf{Objective Results}} \\
\midrule
\textbf{Team} & \textbf{MMSNR} & \textbf{Zimt} & \textbf{FAD} \\
\midrule
xlancelab (ours)      & \textbf{4.4623} & \textbf{0.0137} & \textbf{0.1988} \\
CUPAudioGroup         & 2.3405 & 0.0164 & 0.2253 \\
AC\_DC                & 1.4520 & 0.0182 & 0.2907 \\
Hachimi               & 2.0016 & 0.0183 & 0.2939 \\
cp-jku                & 0.8329 & 0.0189 & 0.3814 \\
\midrule
\multicolumn{4}{c}{\textbf{Subjective Results (MOS)}} \\
\midrule
\textbf{Team} & \textbf{Sep} & \textbf{Rest} & \textbf{Overall} \\
\midrule
xlancelab (ours)      & \textbf{4.2358} & \textbf{3.3892} & \textbf{3.4665} \\
CUPAudioGroup         & 3.8360 & 2.9173 & 2.9253 \\
Hachimi               & 3.5814 & 2.6331 & 2.7235 \\
AC\_DC                & 3.5425 & 2.4768 & 2.5412 \\
cp-jku                & 3.5510 & 2.0838 & 2.1414 \\
\bottomrule
\end{tabular}
\end{table}

\section{Conclusion}
\label{sec:conclusion}

We presented a sequential BS-RoFormer system for the MSR Challenge 2025, achieving top rank. We point out that the current MSR challenge remains primarily focused on separation performance rather than restoration quality. For future challenges, simplifying the task by reducing the number of target instruments could allow participants to focus more resources on restoration quality and advance the field of music source restoration as an individual field of research.

\bibliographystyle{IEEEbib}
\bibliography{refs}

\end{document}